\documentclass{article}
\usepackage{spconf,amsmath,graphicx}
\usepackage{kotex}
\usepackage{xcolor}
\usepackage{url}
\usepackage{mathtools}
\usepackage{graphicx}

\title{Disentangling Timbre and Singing Style with multi-singer Singing Synthesis System}
\name{Juheon Lee, Hyeong-Seok Choi, Junghyun Koo,  Kyogu Lee}
\address{Music and Audio Research Group, Seoul National University \\
\{ juheon2, kekepa15, dg22302, kglee \} @ snu.ac.kr
}
\begin{document}
%
\maketitle
\begin{abstract}
In this study, we define the identity of the singer with two independent concepts -- timbre and singing style -- and propose a multi-singer singing synthesis system that can model them separately. To this end, we extend our single-singer model into a multi-singer model in the following ways: first, we design a singer identity encoder that can adequately reflect the identity of a singer. Second, we use encoded singer identity to condition the two independent decoders that model timbre and singing style, respectively. Through a user study with the listening tests, we experimentally verify that the proposed framework is capable of generating a natural singing voice of high quality while independently controlling the timbre and singing style. Also, by using the method of changing singing styles while fixing the timbre, we suggest that our proposed network can produce a more expressive singing voice. 


\end{abstract}
\section{Introduction}
\label{sec:intro}
Singing voice synthesis (SVS) is a task that generates a natural singing voice from given sheet music and lyrics information. SVS is similar to the text-to-speech (TTS) system in terms of synthesizing natural speech from text information but differs in that it requires controllability of the duration and pitch of each syllable. Similar to the development of TTS \cite{shen2018natural, chen2018sample}, the methodology based on the deep neural network has recently been studied in SVS, and the performance is comparable with the existing concatenative method \cite{blaauw2017neural}.

After the successful development of single-singer model, researches have been conducted to extend the existing model to a multi-singer system. The multi-singer SVS system should not only produce natural pronunciation and pitch contour but also suitably reflect the identity of a particular singer. 
To achieve this, methods for adding conditional inputs reflecting the singer's identity to the network have been proposed \cite{chandna2019wgansing, blaauw2019data}.

In this study, we break down a singer's identity into two independent factors: timbre and singing style.
A timbre is defined as a factor that allows us to distinguish the difference between the two voices even when the singers are singing with the same pitch and pronunciation, and it is generally known that they are related to singers' formant frequency \cite{sundberg2001level, cleveland1977acoustic}. 
Meanwhile, a singing style can be defined as an expression of a singer, hence the natural realization of a pitch sequence from sheet music, including singing skills such as legato, vibrato, and so on.
The expressive SVS system should be able to synthesize the two elements effectively, and it becomes more powerful if the user can control them independently.

To this end, we propose a conditioning method that can model timbre and singing styles, respectively, while extending our existing single-singer SVS system \cite{lee2019adversarially} to a multi-singer system. First, we add a singer identity encoder to the baseline model to capture the singer's global identity. Then we independently condition the encoded singer identity information to the two decoders responsible for formant frequency and pitch contour so that timbre and singing style can be reflected as shown in Fig. 1. Our proposed network can independently control the two identities we define, so cross-generation combining different speakers' timbre and singing styles is also possible. Using this, we generated a singing voice that reflects the timbre or singing style of a particular singer and conducted a listening test, confirming that the network can generate a high-quality singing voice while actually reflecting each identity. 



\begin{figure}[t]
\centering
\includegraphics[width=0.97\linewidth]{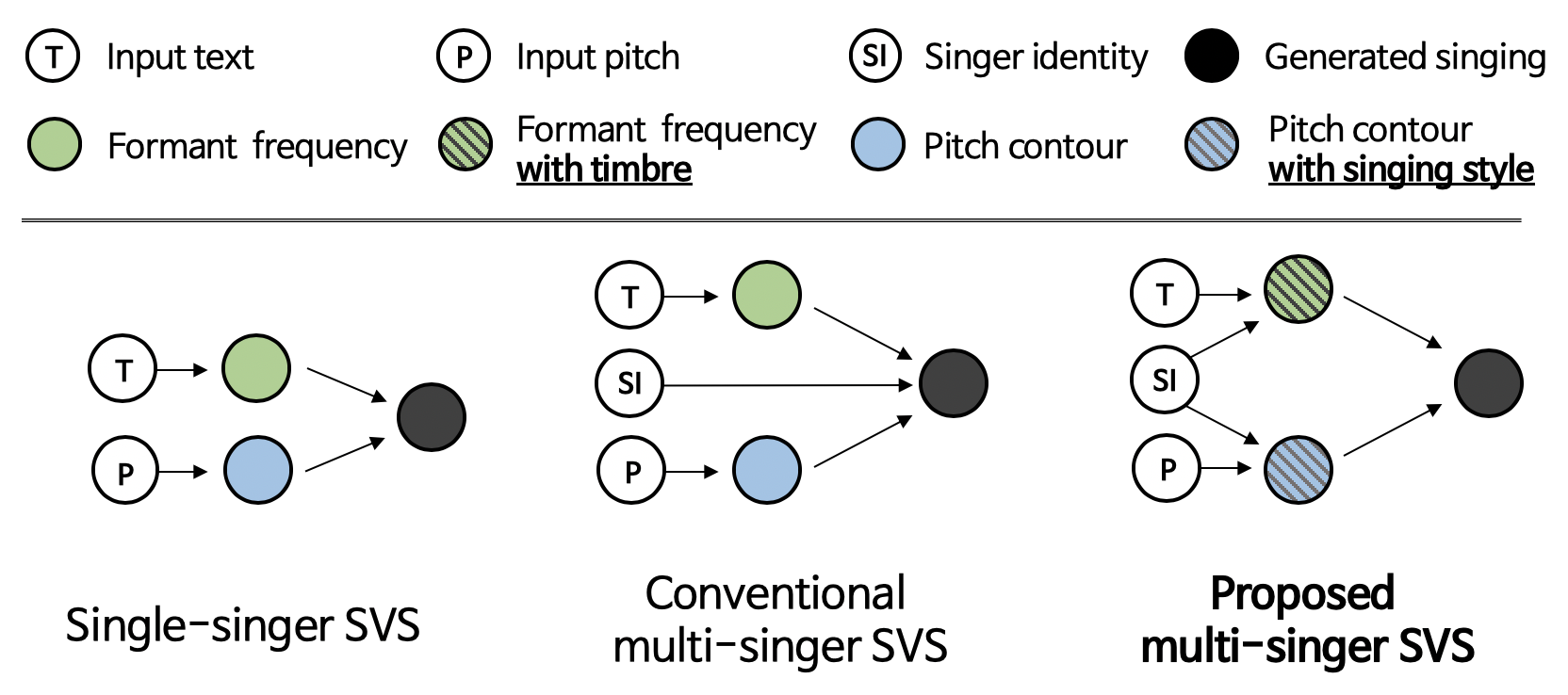}
\vspace{-0.5cm}
\caption{Proposed method to reflect singer identity in multi-singer SVS system} 
\label{fig:proposed}
\end{figure}

The contribution of this paper is as follows: \textbf{1)} We propose a multi-singer SVS system that produces a natural singing voice. \textbf{2)} We propose a new perspective on the identity of the singer -- timbre and singing style -- and propose an independent conditioning method that could model it effectively.

\begin{figure*}[t]
\centering
\includegraphics[width=1\linewidth]{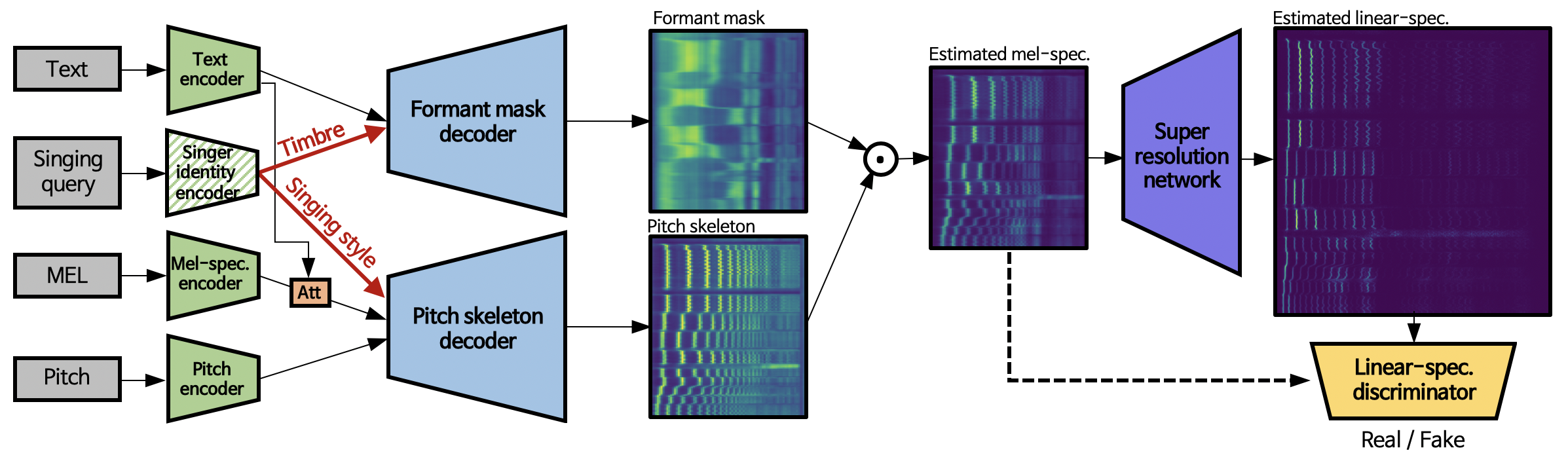}
\vspace{-0.5cm}
\caption{The overview of the proposed multi-singer SVS system.} 
\label{fig:overview}
\end{figure*}

\section{Related works}
\label{sec:related}
\subsection{Single-singer SVS system}
The concatenative method, one of the typical SVS systems such as \cite{macon1997concatenation, kenmochi2007vocaloid, bonada2003sample}, synthesizes the singing voice for a given query based on the pre-recorded actual singing data. This method has the advantage of high sound quality because it uses the human voice directly, but it has a limitation in that it requires an extensive data set every time a new system is designed.
For a more flexible system, parametric methods have been proposed that directly predict the parameters that make up the singing voice \cite{saino2006hmm, nakamura2014hmm, blaauw2017neural}. 
This method overcomes the disadvantages of the concatenative method but has a limitation that depends on the performance of the vocoder itself. Recently, researches are being conducted to generate spectrograms using fully end-to-end methods directly \cite{lee2019adversarially}, or designs of vocoders as trainable neural networks are also in progress \cite{blaauw2019data}. In this study, we experimented based on the end-to-end network that directly generates a linear spectrogram.

\subsection{Multi-singer SVS system}
Researches to extend the SVS system to the multi-singer system has been conducted relatively recently. \cite{chandna2019wgansing} proposes a method of expressing each singer's identity by one-hot embedding. This method is straightforward and simple, but has the limitation of requiring re-training each time to add a new singer. A method of learning trainable embedding directly from the singer's singing query for a more general singer identity is proposed in [2]. 
Our proposed method is different from the previous works in that it directly maps the singing query into an embedding, and defines the singer identity as two independent factors, timbre and singing style.


\section{Proposed System}
\label{sec:pagestyle}

We propose a multi-singer SVS system that can model timbre and singing styles independently. 
We designed the network with \cite{lee2019adversarially} as the baseline and extended the existing model to the multi-singer model by adding 1) singer identity encoder and 2) timbre/singing style conditioning method. 
As shown in Fig. 2, our model uses text $T_{1:L}$, pitch $P_{1:L}$, mel-spectrograms $M_{0:L-1}$, and a singing voice query $Q$ as inputs. Each input is encoded via an encoder, then are decoded with formant mask decoder and pitch skeleton decoder. The formant mask decoder generates a pronunciation and timbre-related feature $FM$ from encoded text $E_T$ and query $E_Q$. The pitch skeleton decoder generates pitch and style-related feature $PS$ from encoded mel-spectrogram $E_M$, pitch $E_P$ and query $E_Q$.
Estimated mel-spectrograms $\hat{M}_{1:L}$; the result of element-wise multiplication of $FM$ and $PS$, are converted to estimated linear spectrograms $\hat{S}_{1:L'}$ via a super-resolution network. Finally, to create a linear spectrogram that is more realistic, we applied adversarial training and added a discriminator to this end. Please refer to \cite{lee2019adversarially} for more detailed information on each module of the network. The summary of the generation process of the entire network is as follows:
\begin{equation}
    \hat{S} = SR(\hat{M}) = SR(FM(T, Q) \odot PS(M, P, Q)).
\end{equation}

\subsection{Singer identity encoder}
Expanding the single-singer model to the multi-singer model requires an additional input about singer identity information. 
To achieve this, we designed a singer identity encoder that directly maps the singer's singing voice into an embedding vector. 
The network structure is shown in Fig. 3. A singing query is passed to two 1d-convolutional layers and an average time pooling layer to capture global time-invariant characteristics while eliminating the changes over time.
Then, the pooled embedding is converted into a 256-dimensional embedding vector through the dense layer and tiled to match the number of time frames of the features. 
Finally, it is used as a conditioning embedding vector for a pitch skeleton decoder and a formant mask decoder, respectively.

\begin{figure}[t]
\centering
\includegraphics[width=1\linewidth]{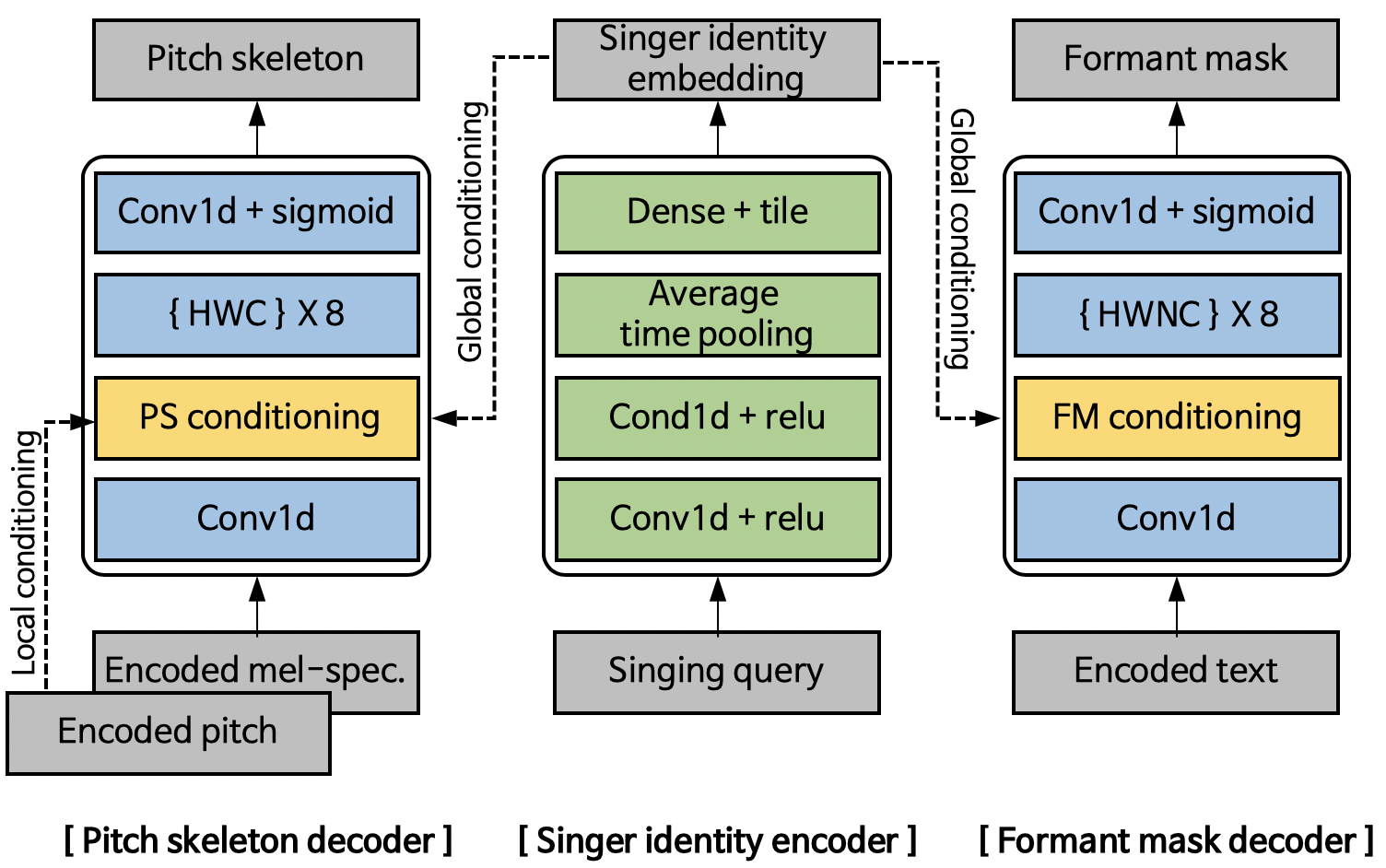}
\vspace{-0.5cm}
\caption{Singer identity encoder structure and conditioning method. HWC, HWNC denotes highwav causal/non-causal covolutional module proposed in \cite{tachibana2018efficiently}, and  $\mathbf{Conv1d, Dense, relu, sigmoid}$ denotes 1d-convolutional layer, fully connected layer, rectifier linear unit, and sigmoid activation unit, respectively.} 
\label{fig:singer_identity_encoder}
\end{figure}

\subsection{Disentangling timbre \& singing style}

In this section, we will provide details of our conditioning method to model timbre and singing styles separately. Our baseline network generates a mel-spectrogram by the multiplication of two different features, formant mask and pitch skeleton. Formant mask is responsible for regulating formant frequency to model corresponding pronunciation information from the input text, while pitch skeleton plays a role in creating natural pitch contours from input pitch. We focused that singer identity embedding could be reflected in each of these features in different ways. In other words, we assumed that singer identity embedding had to be conditioned on the formant mask decoder to control the modality of the timbre, and to control the singing style, it had to be conditioned on the pitch skeleton decoder that forms the shape of the pitch contour. Based on this assumption, we used a method of conditioning singer identity embedding independently of each of the two decoders. We used the global conditioning method proposed in \cite{oord2016wavenet}, and the specific formula is as follows.

\begin{equation}
    \mathbf{z(x, c)} = \sigma(W_1 * \mathbf{x} + V_1 * \mathbf{c}) \odot relu(W_2*\mathbf{x} + V_2* \mathbf{c})
\end{equation}
where $\mathbf{x}$ is a target to be conditioned, $\mathbf{c}$ is a condition vector, and $W_{*} * \mathbf{x} $ and $V_{*} * \mathbf{c} $ are 1d-convolution.

\section{Experiment}
\label{sec:experiment}

\subsection{Dataset and preprocessing}
For training, we use 255 songs of a singing voice, consisting of a total of 15 singers. Three inputs (text, pitch, and mel-spectrogram) were extracted from the lyrics text, midi, and audio data, respectively. Query singing voice for singer identity embedding was randomly chosen from other singing sources of that singer. One of each singer's recorded songs was used as test data, and the rest were used to train the network.

We preprocessed the training data in the same way as \cite{lee2019adversarially} for all input features except the singing query for singer identity embedding. 
The sampling rate was set to 22,050Hz.
The preprocessing step for the singing query is as follows.
First, we randomly selected about the 12-second section from the singer's singing voice source.
Then, we set both the window size and hop length to 1024 and converted the singing voice waveform into a mel-spectrogram of 80 dimensions and 256 frames and used it as the singing query.


\subsection{Training \& inference}
We trained the network in the same way as proposed by \cite{lee2019adversarially}, except to set different speaker samples evenly distributed in each mini-batch. The inference was also conducted in the same way as in the previous study, but for tests to show that the timbre and singing style can be controlled separately, we generated test samples through cross-generation, which generates a pitch skeleton and a formant mask from different speaker embeddings, respectively \footnote{audio samples available at \url{https://juheo.github.io/DTS}}.

\subsection{Analysis on generated spectrogram}


We compared the generated spectrogram by a different speaker for the same pitch and text to see the effect of the speaker identity embedding.
As shown in Fig. 4, each spectrogram has a similar overall shape but includes partial differences. In the case of a formant mask, female vocals have vigorous intensity in high-frequency areas, while the male's corresponding frequency area is shifting down. This is in line with the fact that males generally have lower formant frequency even in the same-pitched condition. Even with the same gender, we can see that the shape of the formant mask is different, and from this, we have confirmed that the speaker embedding appropriately reflects the timbre of each singer. Likewise, pitch skeleton differs depending on the speaker, where it is spotted at the position of the onset/offset, the slope near it, the intensity of vibrato, and the shape of the unvoiced area.
From this, we confirm that the singer identity embedding affects the style change of pitch skeleton effectively. Note that despite conditioning with identical embeddings through time, changes in the style of pitch skeletons over time have been observed. We argue that our network generates singing voice in an auto-regressive way so that it could reflect the style differences over the time axis of different singers.

\begin{figure}[t]
\centering
\includegraphics[width=0.9\linewidth]{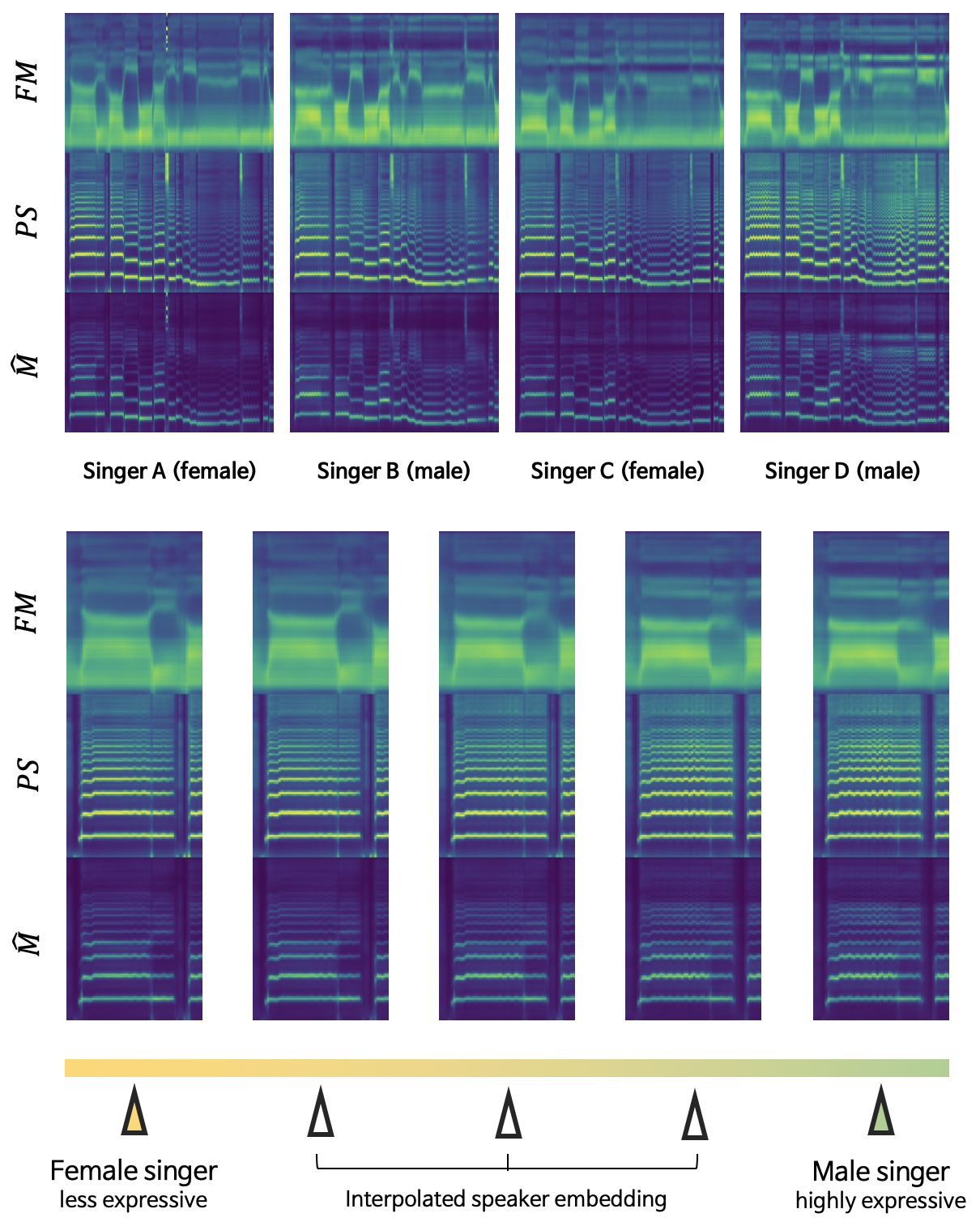}
\caption{Generated mel-spectrogram with various singer embedding (top) and interpolated singer embedding (bottom). $FM$, $PS$, $\hat{M}$ denotes formant mask, pitch skeleton, and estimated mel-spectrogram, respectively.} 
\label{fig:interpolation}
\end{figure}


We were also able to observe a few changes as we interpolate two different singer identity embeddings from female to male vocalist. For example, we found that the high-frequency area of the formant mask was gradually lowered, and the vibrato was gradually strengthened in the case of pitch skeleton. From this, we confirmed that speaker embedding not only reflects the identity of different singers but also contains appropriate information about their changes.
 

\subsection{Listening test}
We conducted a listening test with a total of 6 different male and female singer's voices for qualitative evaluation. We generated two vocal voices for each person for the randomly selected song. To show that the proposed network does not have any degradation in performance even when it independently controls singing style and timbre, we also created two samples for each person's formant mask with another person's pitch skeleton and used them for evaluation. 26 participants were asked to evaluate pronunciation accuracy, sound quality, and the naturalness of test samples. The result is shown in Table 1.

\begin{table}[h]

\caption{ Listening test result (9-point scale) }

\centering

\scalebox{0.8}{

\setlength\tabcolsep{7pt}
\begin{tabular}{c|ccc}

\textbf{Model}      & \textbf{Pronun.acc} & \textbf{Sound.quality} & \textbf{Naturalness} \\ \hline

proposed (w/o cross)     & $7.30 \pm 1.44$ & $5.06 \pm 1.44$ & $5.64 \pm 2.01$ \\
proposed (w/ cross) & $7.36 \pm 1.39$ & $5.19 \pm 1.76$ & $5.55 \pm 2.02$ \\ \hline

Ground     & $7.43 \pm 1.50$ & $6.40 \pm 1.96$ & $6.89 \pm 1.89$ \\

\end{tabular}

}

\end{table}

A paired t-test \cite{ruxton2006unequal} shows no difference for all items, regardless of whether the cross-generation was carried out. We also confirmed that there is no significant difference with the ground truth samples for the pronunciation accuracy. From this, we verify that our proposed network could combine different timbre and singing style without any performance degradation, and can generate a singing voice that can match the ground truth sample with accurate pronunciation.

\subsection{Timbre \& style classification test}

We conducted a classification test to ensure that the network generates results that reflect timbre and singing styles independently. We prepared a total of 20 test sets, 10 each for judging timbre and singing style, and each test set consisted of three sources A, B, and C. A and B are the singing voices generated without cross-generation, and C is cross-generated using its own timbre/style and referencing one of A or B’s style/timbre. By comparing these samples, participants are asked to prefer instead sample C’s timbre/style is a closer match to A or B’s. Considering gender differences, we equally divided three singers’ gender into every possible combination, and the result is as follows in Fig. 5.

\begin{figure}[h]
\centering
\includegraphics[width=1\linewidth]{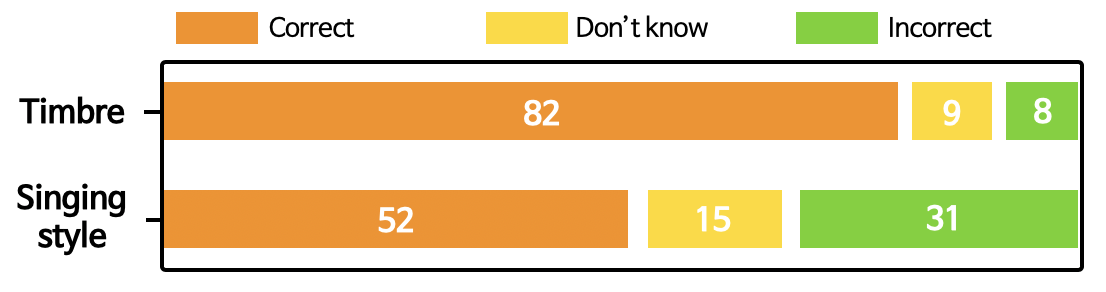}
\vspace{-1cm}
\caption{Timbre and style classification test result} 
\label{fig:timbre_style}
\end{figure}

According to the results of the experiment, 8\% of participants in timbre and 31\% in singing style chose incorrect answers. 
The answer rate of the singing style was lower than the timbre, which is analyzed because the data we used in training consisted of amateur vocals whose style was relatively unclear. Nevertheless, more than half of the participants responded to the correct answer from which we conjecture that our network is able to generate a timbre and singing style that matches a given singer identity query to a level that humans can perceive.


\section{conclusion}
\label{sec:conclusion}
In this study, we proposed a multi-singer SVS system that can independently model and control the singer's timbre and singing style. We disentangled the identity of the singer through a method of conforming singer identity embedding independently in two decoders. The listening test showed that our system could produce high quality and accurate singing comparable to the ground truth singing voice. Through listening tests, which classify the timbre and singing styles of the generated samples, we revealed that we could control both elements independently.




\vfill\pagebreak



\bibliographystyle{IEEEbib}
\bibliography{strings,refs}

\end{document}